\documentclass{aa}
\usepackage{graphicx}
\usepackage{natbib}
\def\ROSAT{{\it ROSAT~~}}
\begin{document}
   \title{The harmonic power spectrum of the soft X-ray\\ background I. The data analysis}

   \author{W. \'Sliwa \inst{1}, A. M. So\l tan \inst{1} \and M. J. Freyberg
\inst{2}}

   \offprints{W. \'Sliwa}

   \institute{Nicolaus Copernicus Astronomical Center, Bartycka 18,
               00-716 Warsaw, Poland\\
              \email{sliwa@camk.edu.pl} (W\'S)\\
              \email{soltan@camk.edu.pl} (AMS)
         \and
             MPI f\"ur extraterrestrische Physik, Giessenbachstr.,
               85748 Garching, Germany\\
              \email {mjf@mpe.mpg.de} (MJF)
             }

   \date{Received   \hspace{20mm};}

  \abstract{
Fluctuations of the soft X-ray background are investigated using harmonic
analysis. A section of the \ROSAT All-Sky Survey around the north galactic pole
is used. The flux distribution is expanded into a set of harmonic functions
and the power spectrum is determined. Several subsamples of the RASS have been used
and the spectra for different regions and energies are presented.
The effects of the data binning in pixels are assessed and taken into account.
The spectra of the analyzed samples reflect both small scale effects generated by strong
discrete sources and the large scale gradients of the XRB distribution.
Our results show that the power spectrum technique can be effectively used to
investigate anisotropy of the XRB at various scales. This statistics
will become a useful tool in the investigation of various XRB components.
   }
   \authorrunning{W. \'Sliwa \& A. M. So\l tan}
   \titlerunning{The XRB harmonic power spectrum. I}
   \maketitle

\section{Introduction}

The diffuse X-ray background (XRB) detected 38 years ago by \cite{Giacc}
was one of the first discoveries in the history of the X-ray astronomy.  
In the seventies, the X-ray satellites {\em ARIEL V} and 
{\em HEAO-1} scanned most of the sky. A large degree of isotropy 
of the X-ray flux found in those investigations
suggested that the origin of the XRB was mainly 
extragalactic and might therefore be of cosmological interest.

At soft energies (below $\sim 4$\,keV) imaging optics of X-ray
telescopes made possible the analysis of the X-ray sky with high sensitivity
and good angular resolution. Observations made with the {\it EINSTEIN}
satellite have shown that a substantial fraction of the soft X-ray background
originates in discrete sources, mostly AGNs (\citealt{giacconi79},
\citealt{tananbaum79}). A further major step in the investigation of the soft XRB
was done using the \ROSAT X-ray telescope (\citealt{truemper90}). Deep \ROSAT
pointing observations resolved $\sim 80$\,\%
of the XRB at $1$\,keV (\citealt{Has1}, \citealt{zrodla}). Using the
\ROSAT observations  \cite{hasinger92} has shown that  at energies $\la 1$\,keV
thermal emission by hot Galactic plasma represents a distinct diffuse
constituent of the XRB.

It is likely that the contribution
to the XRB by the plasma emission is not limited just to our Galaxy.
\cite{cen99} and \cite{phillips00} argue that a non-negligible fraction
of the soft ($0.5-2$\,keV) XRB flux is generated by hot gas in filaments
and halos surrounding galaxies and various galaxy structures.
Thus, the total soft X-ray flux is a mixture of extragalactic 
radiation and emission from the Galaxy. The extragalactic part itself
is non-homogeneous. Apart from the dominating fraction produced by
various classes of X-ray sources, some truly diffuse component 
is expected. In order to study individual components of the XRB it is
necessary to disentangle them carefully from the total flux. 

Investigation of the X-ray background could provide valuable information
on the  Large-Scale Structure of the Universe (\citealt{Bar}).  The observed
XRB flux represents the integrated  X-ray emission along the line of sight.
The bulk of the resolved soft XRB originates at redshifts of $1 - 2$ 
(\citealt{Bar}). It is likely that in the same epoch the largest present-day 
structures collapsed. Since different cosmological models predict a different 
description of the structure formation, the XRB could provide strong
constraints on these models.

One of the basic characteristics of the XRB is the angular distribution
of the background flux in the sky. Measurements of the anisotropy of the XRB
at various angular scales are an effective tool in the study of the
different XRB components. In the present paper we investigate the surface
distribution of the XRB using spherical harmonic analysis. The harmonic power
spectrum is one of the basic statistical instruments in various problems of 
extragalactic astrophysics. It has not been widely used in the studies of the XRB 
because it requires high quality observational material. Recently \cite{scharf}
investigated the all-sky HEAO1-A2 data in the 2-10 keV band in order to
estimate the  XRB dipole. Due to the low resolution of the HEAO\,1 data, the
angular power spectrum of the XRB up to l=20 was calculated. 
The statistical properties of the XRB distribution
have been analyzed in the past using various correlation methods.
Fluctuations of the XRB were successfully measured using the autocorrelation
function (\citealt{soltan9699}). The relationships between different
classes of extragalactic object and the XRB were investigated using 
cross-correlation techniques. This method proved to be very effective in
the analysis of relations between the XRB and normal galaxies (e.g.
\citealt{lahav93}, \citealt{roche95}, \citealt{soltan97}), clusters
of galaxies (\citealt{soltan96}), IRAS galaxies (\citealt{miyaji94}),
and the cosmic microwave background (\citealt{kneissl97}).

The harmonic power spectrum and the autocorrelation function are related
(Eqs.~(\ref{acffin}) and (\ref{mod_a})). Both statistics provide basically
the same information on the anisotropy of the observed distribution.
However, in the actual applications to the specific problems both methods
give results which are complementary, particularly
if the data are subject to large uncertainties. 

The autocorrelation function (ACF) of the XRB has been measured for separations
between $0.^\circ4$ and $\sim 6^\circ$ (\citealt{soltan9699}) using the \ROSAT
All-Sky Survey (RASS). 
Our present measurements of the power spectrum of the XRB are based also
on the RASS. A comparison of the results obtained by these two methods should give
a better insight into the nature of the XRB fluctuations as well as a better
understanding of the errors affecting the RASS data.
To determine the correlations at smaller angular separations
the \ROSAT pointed observations should be used (Soltan et al., in preparation).

In the paper we collect the  formulae of harmonic analysis and give a comprehensive
description of the procedures used in our computations. We calculate the power
spectra of the XRB distribution for several regions of the sky and for different
energy bands. In the future we plan to use these results to investigate the nature
of the XRB fluctuations. The amplitudes and characteristic scales of the XRB
variations determined by means of harmonic analysis will be used to evaluate
various models of the XRB and to specify the contribution of the different components to
the total background flux.

The organization of the paper is as follows. In Sect.~2 we describe
briefly the observational material: the RASS. Because the RASS has
been described in detail in a series of papers (see the references in the
section below), we give here just the basic information. In Sect.~3 the spherical
harmonic analysis is described in detail. We list in a consistent way
all the formulae and describe the modification of the analysis resulting from the
partial sky coverage. A method to verify the accuracy of our computations is also
given. A summary of our results is presented in Sect.~4. We conclude our
investigations with some remarks concerning future work where we plan
to discuss the implications of the present results for various models of the XRB
(Sect.~5).

\section{The data}

The RASS has allowed us for the first time to study an unbiased, spatially
complete sample of the X-ray background.

During the six month survey \ROSAT, using the Position Sensitive Proportional Counter 
(PSPC), scanned the sky in great circles including the ecliptic poles, collecting data
within the $0.1-2.5$\,keV band. The plane of the great circles precessed by
$\sim 4^\prime$ each orbit, following the average solar motion.
Since the full field of view of the PSPC had radius of $57^\prime$, areas of the sky
close to the ecliptic plane were visible for at least 30
orbits or $\sim 2$ days, while the ecliptic poles were covered
by observations during the entire survey.
The exposure therefore increases significantly with ecliptic
latitude, from a minimum of $\sim 700$\,s at the ecliptic plane to over
$50$\,ks at the poles. This variable exposure introduces strong variations
in the signal--to--noise ratio. For a fuller description of the RASS
see \cite{snowden90} and \cite{voges92}.

Various effects and components contaminating the cosmic signal contribute
to the raw data. 
Non-cosmic counts of the  PSPC observations consist mostly of charged
particles and gamma rays which penetrate the detector, small pulse-height events that
follow larger events in the detector and are believed to be  caused  by
negative ion formation and scattered solar X-ray background produced by
scattering of solar X-rays in the Earth's atmosphere. Other sources of non--cosmic
background are auroral X-rays and other sporadic events, most of which are probably 
also produced by low--energy charged--particle interactions with the atmosphere or 
telescope. They  last typically
for a few minutes to a few hours and are called short-term enhancements.
The last component of non cosmic contamination is a long-term enhancement
-- X-ray background of unknown character, that varies on a time scale
of few days and appears most strongly in the R1 and R2 bands.
For details see \cite{snowden94}.

Great efforts were made to separate all these non--cosmic counts from the
genuine X-ray background (\citealt{plucinsky93}, \citealt{snowden94},
\citealt{snowden97}, and references therein). The final RASS maps constitute 
unique material on the large--scale distribution of the XRB.
However, we are interested in the subtle effects which are highly
sensitive to even small imperfections in the data. Thus,
in spite of the careful elimination of the non--cosmic counts,
effects of any potential residual contamination could have 
significant impact on our analysis.

Since most of the non--cosmic background occurs in the low and medium \ROSAT
energy bands, typically below $1$\,keV (with the exception of the particle
contribution which becomes noticeable above $2$\,keV), we used the hard
part of the RASS data, divided into three energy bands:  R5, R6 and R7,
centered at $0.83$, $1.15$ and $1.55$\,keV, respectively. Due to the moderate
spectral resolution of the PSPC, all bands, particularly
the neighboring ones, overlap considerably. Because of this, the
count rates in our bands are highly correlated.

The energy band R6 ($0.73$-$1.56$\,keV) is regarded as the best probe for the
diffuse cosmological XRB, because in this band the systematic uncertainty
induced by the foreground is minimized. We concentrate
our analysis and results on R6 band, but investigate systematic effects
by comparison with the two neighboring bands, particularly the R5 band.
This band is also low in non-cosmic photons and in
contamination by charged particles, but, as the amplitude of the galactic
component relative to the extragalactic signal increases drastically towards
soft energies, the R5 band contains an increased galactic
foreground component compared to the R6 band. Thus the R5 band should allow
us to discriminate between the galactic and the extragalactic signals.

\begin{figure}
\centering
\includegraphics[width=0.8\linewidth]{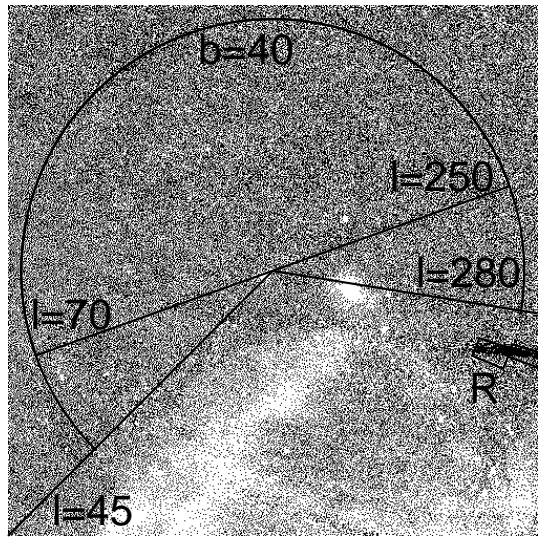}
\caption{Section of the RASS centered on the North Galactic Pole in
an equal area projection. The whole map covers $3.19$ Sr. The areas defined
in Table~1 used in the computations are marked. The small region marked `R' has been removed
from the data because of insufficient exposure time. \label{map}}
\end{figure}

The data from the north galactic hemisphere are used. The XRB flux is binned
into $12^\prime\times 12^\prime$ pixels composing an array of $512\times 512$
elements. An equal-area polar projection centered on the North Galactic Pole
has been applied to record the data. Within this area a small patch\footnote
{For the present computations this region is defined as ($287^\circ < l < 290^\circ,\;
30^\circ < b < 40^\circ$) and ($288^\circ < l < 293^\circ,\;40^\circ < b < 47^\circ$).}
of sky marked `R' in Fig.~\ref{map} has not been covered
by the RASS. A large portion of the north galactic hemisphere is dominated
by the hot plasma emission in our Galaxy (\citealt{snowden95}).
Most of this local signal is concentrated
at galactic longitudes between $260^{\circ}$ and $50^{\circ}$. The remaining
area is visually free from the foreground galactic signal, but some local
contribution is also present (\citealt{soltan9699}). To investigate effects
of the galactic contamination on our determinations of the power spectrum,
computations have been performed using various subsets of the available data.

The regions investigated in detail in the present analysis are defined in Table~1.
The total area (`A' in Table ~1) is highly non-homogeneous because of the
galactic component. The spherical harmonic transform for this area has
been computed mainly as a check of our algorithm (see below). Area `B'
is the largest fragment of the northern celestial hemisphere apparently
free from strong galactic emission. However, the border regions of this area
are close to prominent emission regions. Area `B' extends also to low
galactic latitudes ($b < 20^\circ$ at the ''corners" of the RASS map) and
effects of the galactic obscuration are likely to be present. Thus, one could
expect some interference of galactic and extragalactic signals. Area `C'
is restricted to the galactic latitudes $b > 40^\circ$. The smallest area , `D',
has been extensively analyzed by \cite{soltan9699}. 
The power spectrum obtained for this region should correspond to the
autocorrelation function obtained in those papers. To see the effects on our
results of the brightest sources, some computations were performed without
two conspicuous clusters: Coma and Abell~1367.

\begin{table}
\begin{center}
\begin{tabular}{|c|r|r|c|r|r|} \hline \cline{1-5}
\multicolumn{5}{|c|}{\bf Table 1 }\\
\hline \cline{1-5}
Sample&$l_{min}$&$l_{max}$&$b_{min}$&$\Omega$[Sr]\\
\hline
  A   &   $ 0^\circ$ & $360^\circ$  &$12^\circ - 37^{\circ\dagger}$& $3.18^\ddagger$   \\
\hline
  B   &   $45^\circ$ & $280^\circ$  &$12^\circ - 37^{\circ\dagger}$& $2.04$   \\
\hline
  C   &   $45^\circ$ & $280^\circ$  &       $40^\circ$             & $1.43$   \\
\hline
  D   &   $70^\circ$ & $250^\circ$  &       $40^\circ$             & $1.12$   \\
\hline
\end{tabular}

\end{center}
~\\
$^\dagger$~Minimum galactic latitude varies with the galactic longitude
(see Fig.~\ref{map}).\\
$^\ddagger$~After removal of region `R'.
\end{table}

\section{The power spectrum}

Although the definitions and basic relationships for spherical harmonics
are outlined in a number of astrophysical publications, we describe below
in some detail the standard formulae and all the essential steps of our
analysis in order to present it in a complete and coherent way. Where possible, 
we followed the notation used by \cite{peebles73} and \cite{hauser73}.

\subsection{Sky coverage}

Spherical harmonic  functions constitute a set of orthogonal functions on the sphere.
Astrophysical data rarely cover a full solid angle of $4\pi$. In the present
investigation the useful data are also available for a limited region of the sky
only. Incomplete sky coverage destroys the orthogonality of the spherical harmonic functions.
Consequently, the power spectrum coefficients are correlated
(\citealt{hauser73}). There are basically two ways to take this effect into
account. One can either construct a new set of functions which are orthogonal in the 
defined area of the sphere (e.g. \citealt{gorski94},~\citealt{tegmark96}), or fill the area 
where the observational material is missing with synthetic data (see below). Both
methods have their advantages as well as drawbacks. In the present paper we use the latter
approach for two reasons. Firstly, we perform calculations using data selected in several
regions of the sky. Different windows imply different sets of orthogonal functions,
which makes comparison of the results ambiguous. Secondly, we are interested in XRB
fluctuations at angular scales significantly smaller than the size of the area covered
by the data. A section of the harmonic power spectrum corresponding to these small scales
is adequately determined using the available RASS data.

\subsection{Spherical harmonics - the formulae}

The distribution of the XRB flux on the celestial sphere, $\rho(\theta,\phi)$,
is represented as an infinite sum of spherical harmonics $Y_l^m$:

\begin{equation}
\rho(\theta,\phi) = \sum_{l=0}^{\infty}\,\sum_{m=-l}^l\;
a_l^m\,Y_l^m(\theta,\phi),                                             \label{suma}
\end{equation}
where the coefficients $a_l^m$ are the spherical harmonic transform: 
\begin{equation}
a_l^m = \int_{4\pi}\,\rho(\theta,\phi)\:Y_l^{m\ast}(\theta,\phi)\:d\Omega\,.
                                                                       \label{amldef}
\end{equation}

The RASS covers practically the whole celestial sphere, but only
relatively small regions of the sky contain the data dominated by the
extragalactic signal; the remaining areas are heavily contaminated by 
local effects. As a result of this, we are forced to carry out the power
spectrum analysis using the data from a fraction of the sky.  To assess
the effects of incomplete sky coverage we introduce, following \cite{peebles73},
the window function, $W(\theta,\phi)$, which describes the area used in the
analysis:

\begin{eqnarray} \label{windef}
W(\theta,\phi) = \cases{ 1, & \hbox{inside the accepted region,}\cr
                 0, & \hbox{outside the accepted region.}}
\end{eqnarray}
Using the window function we define for the whole celestial sphere a model
distribution of the XRB:

\begin{equation}
\tilde{\rho}(\theta,\phi) = W(\theta,\phi)\,\rho(\theta,\phi) + 
    [1-W(\theta,\phi)]\,\langle\rho\rangle\,,                     \label{modeldef}
\end{equation}
where $\langle\rho\rangle$ is the mean background flux averaged over
the accepted area. Thus, within the window area the model distribution is
identical 
to the actual one, while outside the window it is constant and equal to the
average signal computed using the data just from the accepted region.
The model distribution covers the whole sphere, but the cosmic
information is limited to the window area.

The spherical harmonic transform of $\tilde{\rho}$,
\begin{equation}
\tilde{a}_l^m = \int_{4\pi}\,\tilde{\rho}\:Y_l^{m\ast}\,d\Omega\,,     \label{tilamldef}
\end{equation}
and $\tilde{\rho}$ is related to the $\tilde{a}$'s by a formula
analogous to Eq.~(\ref{suma}). Using Eq.~(\ref{modeldef}) we get:
\begin{eqnarray}
\label{cmlres}
  \tilde{a}_l^m = \cases{\int_{\Omega}\,\rho\:Y_l^{m\ast}\:d\Omega -
                \langle\rho\rangle\,\int_{\Omega}\,Y_l^{m\ast}\,d\Omega,
                                    & \hbox{for $l\neq 0$,}\cr
      \sqrt{4\pi}\,\langle\rho\rangle, & \hbox{for $l = 0$,}}
\end{eqnarray}
where the integrals extend over the solid angle, $\Omega$, of the accepted area.
The zero degree term, $\tilde{a}_0^0$, represents the average signal in the data and
does not depend on the window function. One should note that coefficients
$\tilde{a}_l^m$, although calculated using the cosmic data from the fraction of the
sky, represent a mathematically well defined spherical harmonic transform. In
particular, the expected values of all the  $l \neq 0$ coefficients
$\tilde{a}_l^m$ are equal to 0. Eq.~(\ref{cmlres}) implies that
\begin{equation}
\left\langle \int_{\Omega}\,\rho\:Y_l^{m\ast}\,d\Omega\right\rangle =
     \langle\rho\rangle \int_{\Omega}\,Y_l^{m\ast}\,d\Omega\,,   \label{renorm}
\end{equation}
where $\langle ... \rangle$ enclosing the integral denote the expectation value.
Eq.~(\ref{renorm}) shows that the `spherical harmonic transform' coefficients
calculated from a fraction of the sphere deviate systematically from zero
(see \cite{peebles73} for a detailed discussion).

One should expect that at angular scales substantially smaller than the
characteristic size of the window, the fluctuations of the XRB are adequately
described by $\tilde{a}_l^m$. In order to find a relationship between
$\tilde{a}_l^m$ and ${a}_l^m$, we use the angular autocorrelation function
(ACF), $w(\theta)$:
\begin{equation}
w(\theta)={{\left\langle\rho(\omega_1)\,\rho(\omega_2)\right\rangle}_{\mid{4\pi}}
\over \langle\rho\rangle^2} - 1\,,
\label{acfdef}
\end{equation}
where $\omega_i = (\theta_i,\phi_i)$ for $i=1,2$ denote points in the
celestial sphere separated by the angle $\theta$ and the average values
$\langle ... \rangle$ are calculated for the whole sphere.
Using Eq.~(B2) of \cite{peebles73} and Eq.~(\ref{suma}) one obtains the relationship
between the average flux correlation and the spherical harmonic transform:
\begin{equation}
{\left\langle\rho(\omega_1)\,\rho(\omega_2)\right\rangle}_{\mid{4\pi}} =
   \sum_{l=0}\;\sum_{m=-l}^l \mid\!a_l^m\!\mid^2\,{P_l(\cos\theta)\over 4\pi}
                               \label{corrho}
\end{equation}
\begin{displaymath}
\phantom{{\left\langle\rho(\omega_1)\,\rho(\omega_2)\right\rangle}_{\mid{4\pi}}}
    =\langle\rho\rangle^2 + \sum_{l=1} {2l+1\over 4\pi} Z_l\,P_l(\cos\theta)\,,
                        \label{acfps}
\end{displaymath}

where 
\begin{equation}
Z_l = {1\over 2l+1}\,\sum_{m=-l}^l \mid\!a_l^m\!\mid^2   \label{psdef}
\end{equation}
is the power spectrum for the distribution on the sphere, and $P_l$
is a Legendre function of degree $l$. In an analogous way we define the
power spectrum of the model distribution $\tilde{\rho}$:
\begin{equation}
\tilde{Z}_l = {1\over 2l+1}\,\sum_{m=-l}^l \mid\!\tilde{a}_l^m\!\mid^2\,.
                                                         \label{tilpsdef}
\end{equation}
Eqs.~\ref{acfdef}, \ref{corrho} and \ref{psdef} give
\begin{equation}
w(\theta) = {1\over\langle\rho\rangle^2}\;
                 \sum_{l=1} {2l+1\over 4\pi}\,Z_l\,P_l(\cos\theta)\,, \label{acffin}
\end{equation}
and an analogous relationship between $\tilde{w}$ and $\tilde{Z}_l$.
If we now limit our analysis to separation
angles small in comparison to the size of the area of solid angle $\Omega$,
we may express the average product of the model flux
$\langle\tilde{\rho}(\omega_1)\,\tilde{\rho}(\omega_2)\rangle_{\mid4\pi}$
as the sum of two components
(using Eq.~(\ref{modeldef})):
\begin{equation}
\langle\tilde{\rho}(\omega_1)\,\tilde{\rho}(\omega_2)\rangle_{\mid4\pi}\approx
{\Omega\over 4\pi}\,\langle\rho(\omega_1)\,\rho(\omega_2)\rangle_{\mid\Omega} +
{4\pi - \Omega\over 4\pi}\,\langle\rho\rangle^2\,,
\end{equation}
where on the left hand side the averaging is over the whole sphere,
while on the right hand side it is done over a solid angle of $\Omega$. One can transform
this equation into:
\begin{equation}
\tilde{w}(\theta) = {\Omega\over 4\pi}\,w_{\Omega}(\theta)\,, \label{acfrel}
\end{equation}
where $w_{\Omega}$ is the autocorrelation function calculated using the data
within the solid angle $\Omega$. As long as the cosmic signal contained within
the window is a {\it fair sample} of the entire XRB, $w_{\Omega}(\theta) = w(\theta)$.
Thus, scaling of the ACFs (Eq.~(\ref{acfrel})) together with Eqs.~(\ref{psdef}) 
and (\ref{tilpsdef}) gives
\begin{equation}
Z_l = {4\pi\over\Omega}\,{1\over 2l+1}\,\sum_{m=-l}^l\mid\!\tilde{a}_l^m\!\mid^2\,,
                            \label{ps}
\end{equation}
in agreement with Eqs.~(53) and (54) by \cite{peebles73}. Eq.~(\ref{ps}) shows that
the high $l$ part of the power spectrum (for the whole sphere) is directly
proportional to the power spectrum based on the data from the fraction of the sphere
if these data are representative for the whole sphere (fair sample assumption).

\subsection{Spherical harmonics - computations}

\begin{figure}
\centering
\includegraphics[height=0.8\linewidth]{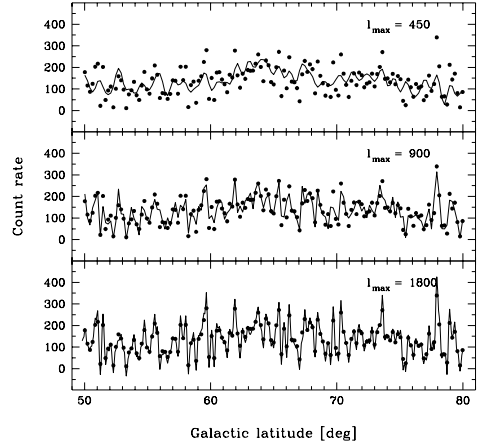}
\caption{Count rate distribution in the R6 energy band of the RASS in $12'$
pixels along galactic longitude $l=0$. The data in this region are heavily
contaminated by galactic emission which causes large-scale variations
in the count rate. The curves show the synthesized distribution for $l_{\rm max} =
450$, {900}, and {1800}. \label{section1}}
\end{figure}

\begin{figure}
\centering
\includegraphics[width=0.8\linewidth]{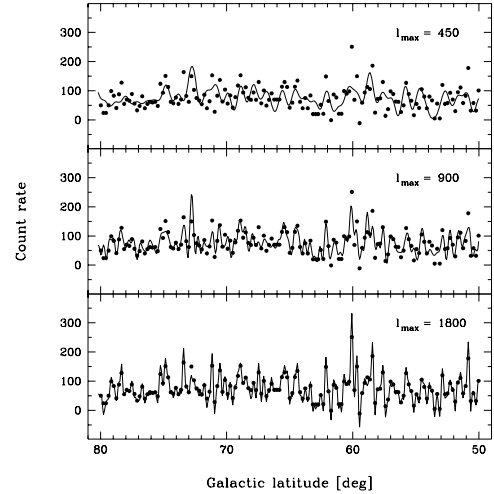}
\caption{Same as Fig.~\ref{section1} for galactic longitude $l = 90^{\circ}$.
Most of the signal is of extragalactic origin. Because of the negligible
galactic contribution, the average count rate is lower than in
Fig.~\ref{section1}. Note the good fit for $l_{\ rm max} = 1800$ (bottom panel).
\label{section2}}
\end{figure}

The large survey area combined with a relatively high angular resolution of
the RASS allows us to calculate the power spectrum over a wide range of 
the spherical harmonic degree $l$. Fluctuations
of the XRB on the scale $\theta$ will produce a signal in the power spectrum
at $l \sim \pi/\theta$. Since the pixel size of the RASS is $12'$,
our algorithm should be effective up to $l \sim 900$. Fortunately
this requirement could be satisfied with the present day computers without
major problems.

The spherical harmonic function is defined in a standard way by:
\begin{equation}
Y_l^m(\theta,\phi) = c_l^m\,P_l^m(\cos\theta)\,e^{im\phi}\,, \label{ylm}
\end{equation}
\begin{displaymath}
c_l^m = 
\left[{2l+1\over 4\pi}\,{(l-m)!\over(l+m)!}\right]^{1/2}\,,
\end{displaymath}
\begin{displaymath}
P_l^{-m}(x)=(-1)^m\,{(l-m)!\over (l+m)!}\,P_l^m(x)\,.
\end{displaymath}
where $P_l^m$ are the associated Legendre polynomials. For positive m
these are defined by the formula:
\begin{equation}
P_l^m(x)=(-1)^m\,(1-x^2)^{m/2}\,{d^m\over dx^m}P_l(x)\,.  \label{plmasc}
\end{equation}
Using the  above relations it can be seen that
\begin{equation}
Y_l^{-m}(\theta,\phi)=(-1)^m\,Y_l^{m\ast}(\theta,\phi)\,.
\end{equation}
In the actual computations it is convenient to use recursive formulae rather
than Eqs.~(\ref{ylm}) and (\ref{plmasc}) directly. We use the relations given by 
\cite{brauthwaite73} as quoted by \cite{hauser73}.
To verify the accuracy of our computations and demonstrate the effects of
the maximum harmonic order on the resolution and quality of the fits,
we synthesized the model distribution
using Eq.~(\ref{suma}) and the spherical harmonic transform
coefficients $\tilde{a}_l^m$. Because of the finite number of spherical
harmonics used in the calculations the synthesized map does not reproduce
exactly the original RASS data. 

In Fig.~\ref{section1} we show a sample distribution of count rates in
146 pixels along the line of galactic longitude $\ell = 0$
between galactic latitudes $50^\circ$ and $80^\circ$
\footnote{Strictly speaking, one side of the pixel row is located exactly
at $\ell = 0$, while the pixel centers are shifted by $6^\prime$.}.
The three panels show the synthesized distributions for three values of the
maximum spherical harmonic degree used in the calculations,
$l_{\rm max} = 450$, $900$, and $1800$. The spherical harmonic transform
$\tilde{a}_l^m$ was calculated for the total area of roughly $3.2$\,sr
shown in Fig.~\ref{map}. In the top panel with $l_{\rm max} = 450$ the synthesized
distribution follows the large scale features of the real distribution. The
minimum angular scale `resolved' in this model corresponds roughly
to 2 RASS pixels. The fit to the real signal looks bad because the model locally
represents the signal averaged over $\sim 4$ pixels and the scan in
Fig.~\ref{section1} has the width of a single pixel. The fit is substantially
improved in the middle panel where $l_{\rm max}=900$ and still further in the
lower panel for $l_{\rm max}=1800$. It is evident that to reproduce details of the 
XRB distribution binned into $12^\prime\times12^\prime$ pixels the harmonic decomposition 
has to reach the degree of $l_{\rm max}\approx900$.
The model with $l_{\rm max}=1800$ reproduces the features of the cosmic signal
very accurately. We treat this as a verification of our computational algorithm.
In Fig.~\ref{section2} the section of the RASS for
galactic longitude $90^{\circ}$ is shown. Here the contamination by the
galactic plasma emission is small and the signal is dominated by the
extragalactic component. The conspicuous large-scale trends are absent
in this region and the fluctuations are dominated by point-like sources.
Similar to Fig.~\ref{section1} the goodness of the fit improves substantially
with increasing $l_{\rm max}$. In particular, the fits for $l_{\rm max} \ga 900$
adequately reproduce the peaks associated with bright sources.

\subsection{Effects of pixelization of the data}

The power spectrum at high $l$ is affected by the pixelization
of the XRB data. The XRB fluctuations at scales smaller than the pixel size
are smoothed over the pixel area modifying the flux correlation at
this angular scale and removing the variations at smaller scales.
We assess these effects as follows.

\subsubsection{Fluctuations below the pixel size \label{correct_below}}

\begin{figure}
\centering
\includegraphics[width=0.8\linewidth]{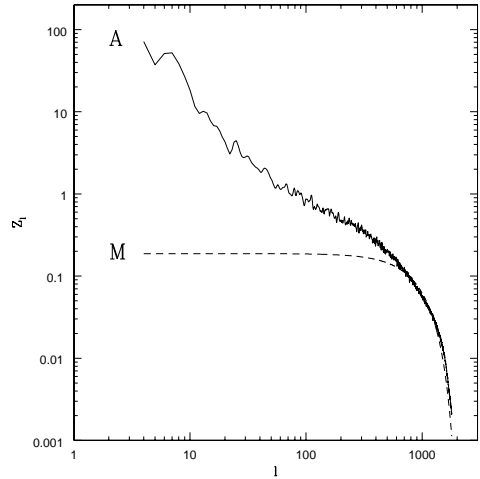}
\caption{A - Power spectrum of all data (Sample A); M - model
which shows the contribution to the overall spectrum by fluctuations
at scales below the pixel size.
\label{ps_full}}
\end{figure}

The count rate fluctuations recorded in the $12'$ RASS pixels result from
the cosmic variations and the photon noise. Cosmic variations are produced
mainly by discrete sources and should be distributed uniformly (apart from
the source clustering) at high galactic latitudes. On the other hand,
the amplitude of the fluctuations generated by the photon statistics
depends on the exposure time and varies over the survey area. In principle,
this property should help to distinguish between these two sources of
fluctuations. We plan to analyze this effect in the future paper. 
Due to the presence of extended X-ray sources and source clustering, the
cosmic signal binned in pixels is correlated, while the fluctuations
generated by the photon noise are uncorrelated. 

Uncorrelated fluctuations generate the flat power spectrum over all $l\la 900$.
To assess the contribution to the power spectrum by fluctuations at scales below
the pixel size, we calculate the model spectrum assuming that the count
rates in pixels are uncorrelated.  We use a formula
analogous to Eq.~(22) of \cite{peebles73}:
\begin{equation}
\langle\mid\!a_l^m\!\mid^2\rangle=2\pi \langle\rho\rangle^2
    \int_{-1}^{+1} w(\theta)\,P_l(\cos\theta)\,d\cos\theta\,, \label{mod_a}
\end{equation}
which is obtained from Eq.~(\ref{amldef}) above and Eq.~(B2) of \cite{peebles73}.
It is straightforward to calculate the exact shape of the autocorrelation
function assuming a `flat' X-ray signal within each pixel and no correlation
between pixels. Such a synthetic autocorrelation function depends only on the
second moment of the count rates in the pixels and on the pixel size (and shape);
this model autocorrelation function is equal to zero for separations which
are not contained within a single pixel. The solid curve in Fig.~\ref{ps_full}
shows the power spectrum $Z_l$ calculated for the total area (Sample A
of Table 1) and the dashed curve represents the model power spectrum, 
assuming that the fluctuations in $12^\prime$ pixels are uncorrelated.
The observed power spectrum is fitted well by the model for $l\ga 700$ and
has a systematically larger amplitude for smaller $l$. This demonstrates
that the cosmic signal is strongly correlated over a wide range of separations.
Additionally, good agreement in the high-$l$ range, where potentially 
numerical problems could develop, indicates that our algorithm is accurate
over the entire range of $l$.

\begin{figure}
\centering
\includegraphics[width=0.8\linewidth]{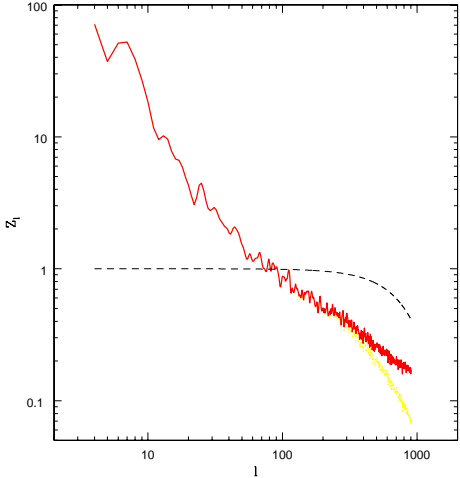}
\caption{The raw power spectrum of all data. Sample A: dotted curve;
the corrected spectrum: solid curve; the correction for the finite
pixel size: dashed curve.
\label{ps_corr}}
\end{figure}

\subsubsection{Corrections for moderate harmonic scales \label{corrections}}

The dashed curve in Fig.~\ref{ps_full} shows that binning the data into
$12^\prime$ pixels reduces the amplitude of the power spectrum also
at degrees below $l=900$. To account for this effect we have calculated
the distortion of the power spectrum generated by the binning. The
model power spectrum $Z_l = 1$ was used as input data. This flat
spectrum represents a random distribution of point sources. Due to
the binning of the input flux, the resulting power spectrum has a
shape analogous to the model marked `M' in Fig.~\ref{ps_full}. The ratio
of the modified power spectrum to the input model is shown in Fig.~\ref{ps_corr}
by a dashed line. To correct for this effect one should normalize
the raw RASS spectra by this curve. 
In Fig.~\ref{ps_corr} the dotted curve represents the raw spectrum of the
XRB in area `A', and the solid curve shows the corrected spectrum. 
The spurious curvature of the spectrum caused by the data binning is removed. This
procedure is effective up to a harmonic degree of $l\approx 900$ and is applied
to all subsequent power spectra computations.

\section{Results}

\begin{figure}
\centering
\includegraphics[width=0.8\linewidth]{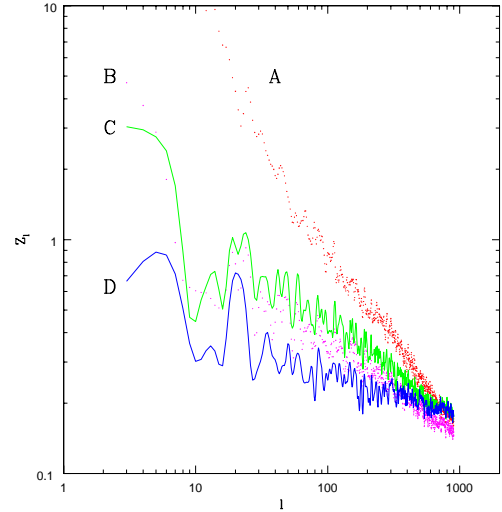}
\caption{Power spectra of the XRB in the R6 band. Labels `A', `B', `C' and `D'
denote the RASS regions defined in Table~1.
\label{ps_r6_abcd}}
\end{figure}

In Fig.~\ref{ps_r6_abcd} the power spectra of the R6
band are shown for four regions of the RASS defined in Table 1.

The spherical transform at low degrees is corrupted by the window function.
The maximum angular scale effectively
accessible to the analysis is several times smaller than the size of the area
used in the computations. The strong oscillations of $Z_l$ for areas
`B' -- `D' visible in Fig.~\ref{ps_r6_abcd} indicate that the power
spectrum estimates of any individual degree $l$ are subject to large
uncertainties at low $l$ values. Assuming that the `true' shape of the
power spectrum is more regular than that in Fig.~\ref{ps_r6_abcd},
one can try to fit a smooth function. This would extend our $Z_l$ estimates
down to $l\approx 10$.

Apart from small $l$ variations none of the spectra exhibits any distinct features
in the entire range of degrees. The lack of significant spectral structures is also 
apparent in Fig.~\ref{ps_r6}. The RASS data have been accumulated in the scanning mode of
the satellite operation.
Thus the raw data consist of a large number of $\sim 2^\circ$ wide strips
along great circles of constant ecliptic longitudes. In a laborious procedure
(\citealt{snowden95} and references therein)
these raw counts have been reduced to create a homogeneous map of the XRB flux.
Despite a great effort to remove all the artifacts produced by the data
accumulation technique, one should be concerned that not all the instrumental effects
have been removed from the final maps. Our power spectra do not show visible
features at degrees corresponding to an angular scale of $2^\circ$, which 
indicates that the XRB maps are free from noticeable effects generated
by the scanning mode.

The conspicuous difference of the power spectrum at low degree $l$ between 
area `A' and the remaining areas results from strong contamination
by galactic emission  between galactic longitudes $280^\circ$ and
$45^\circ$. One should also note  that the power spectrum amplitude for 
area `D' is systematically smaller than for areas `B' and `C'.
This difference indicates that the contribution of the hot galactic
plasma to the background is not confined to the region between
$280^\circ$ and $45^\circ$ but extends further away. It is important
that, apart from differences at small $l$, all power spectra
converge at high $l$, indicating that fluctuations at the smallest scales are
dominated by the extragalactic component. Although the contribution
of the hot plasma to the spectrum at larger scales is substantially
reduced in area `D' in comparison with the remaining areas, it is
still possible that some signal is generated locally. To investigate
this problem, we now concentrate on the analysis of region `D'.

\subsection{Error estimates}

\begin{figure}
\centering
\includegraphics[width=0.8\linewidth]{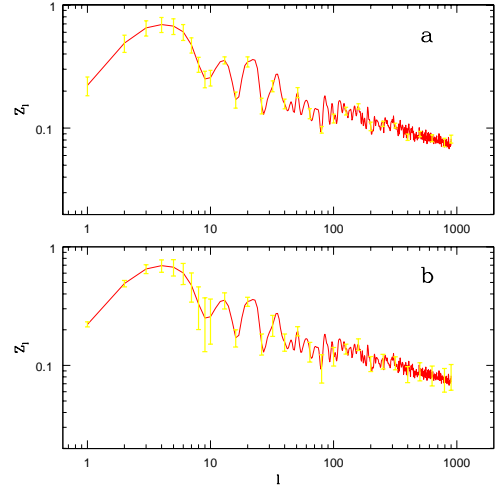}
\caption{Uncertainties of the power spectrum estimates produced by photon
counting statistics (a) and cosmic variance (b). Area `D' is used. 
The error bars represent $1\sigma$ uncertainties. For clarity, the error
bars are plotted for selected degree $l$.}
\label{sigma}
\end{figure}

Our estimates of the power spectrum are subject to two sources
of errors. Limited counting statistics and instrumental background subtraction
produce uncertainties in the distributions of the XRB flux in the
RASS data. Information on the expected rms count rate uncertainty for each
pixel has been obtained  during the RASS production process and is stored in
a separate array. We used these data to generate 24 simulated RASS maps.
For each simulation the count rates in all pixels have been randomized using
a Gaussian distribution with an average equal to the real count rate and
a variance taken from the uncertainty array. The power spectra of all the simulations
have been computed. The rms scatter between these simulated functions is used
as an estimate of the $1 \sigma$ uncertainty of the power spectrum of the real data.
The corresponding error bars are shown in Fig.~\ref{sigma}a for one of
the spectra of area `D'.

A second source of uncertainty is related to the cosmic variance. To account for this
effect, we have divided area `D' into 6 sections and computed the
power spectra for each region separately. The scatter between the spectra of
the 6 regions is used to estimate the uncertainty of the spectrum based on
the entire area `D'. The resultant error bars are shown in Fig.~\ref{sigma}b.
Since the average exposure times in the RASS are rather short, the uncertainties
in the count rate for individual pixels are relatively high. Nevertheless,
one should note that the uncertainties introduced  by cosmic variance are larger
than the errors generated by poor counting statistics.

\subsection{The power spectrum of the R6 band}

\begin{figure}
\centering
\includegraphics[width=0.8\linewidth]{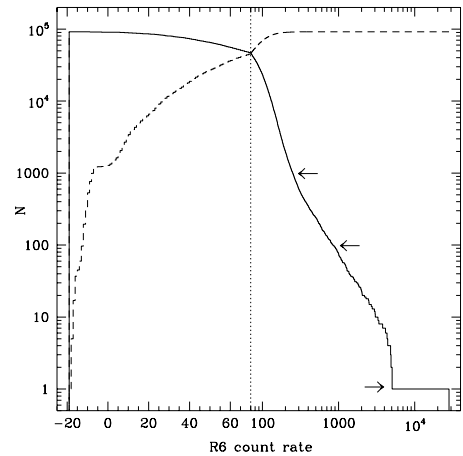}
\caption{The cumulative distribution of the count rates in the pixels
in area `D' in the R6 band. The solid curve gives the number of $12^\prime\times
12^\prime$ pixels with a count rate above the specified value; the dashed curve
gives the complementary  distribution. Linear and log scales are used for low and high 
count rate sections of the distribution. The vertical dotted line (which splits the plot) 
marks the median of the distribution. Arrows indicate the cutoffs used in the calculations
(see text). \label{crt_hst1_r6}}
\end{figure}

\begin{figure}
\centering
\includegraphics[width=0.8\linewidth]{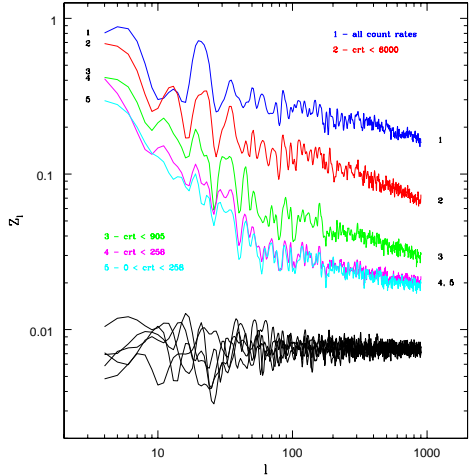}
\caption{The power spectra in area `D' with different thresholds imposed
on the count rates: 1: all data (same as in Fig.~\ref{ps_r6_abcd}); 2:  the pixel with
the highest count rate has been removed; 3: the pixels with the top $0.1$\,\% count rates 
have been removed; 4: the pixels with the top $1.0$\,\% count rates have
been  removed; 5; the pixels with
the top $1.0$\,\% count rates and with negative count rates have been removed;
the curves
in the lower part represent five simulations (see text).
\label{ps_r6}}
\end{figure}

The power spectra shown in Fig.~\ref{ps_r6_abcd} are calculated using
all the XRB flux in the specified area including strong sources. Many of these
sources produce count rates which exceed the average count rate per pixel
by a factor of ten or more. In Fig.~\ref{crt_hst1_r6} we have plotted
the cumulative distribution of the count rates\footnote{In  \ROSAT `working' units 
of $10^{-6}\times$ PSPC counts s$^{-1}$ arc\,min$^{-2}$.} in the R6 band
for area `D'. The distribution has a long high count rate tail extending to above
$28\cdot 10^3$, while the average of the distribution is equal to $\sim 80$.
The highest signal, viz. 28175, in the R6 band is generated by a source associated
with the Seyfert galaxy IC\,3599 (\citealt{komossa99}). In $0.1$\,\% of pixels the count 
rate exceeds $904$ and in $1.0$\,\% it exceeds  $256$.
Such peaks generate a signal in the power spectrum over all harmonic degrees.
To see what impact these strong peaks have on the resulting power spectrum we have
calculated the spectra using the data without the brightest sources. This was
achieved by modifying the window function (Eq.~(\ref{windef})) to exclude pixels
with count rates above the assumed threshold. The effects of the high count rate
tail on the power spectra are shown in Fig.~\ref{ps_r6}. 

Since the narrow  peaks in the distribution produce a signal in the harmonic power
spectrum over the entire range of degrees $l$, the removal of the brightest sources
from the area reduces the amplitude of the power spectrum over a wide range of degrees.
As a result of the  subtraction of the peaks the shape of the spectrum changes significantly.
For the whole data (Fig.~\ref{ps_r6}, spectrum No 1) it is roughly a power law
for $l < 900$, while spectrum No 4 which represents the data without the brightest sources is
well approximated by a broken power law with steep slope at degrees below
$\sim 60$ and distinctly flatter slope at larger $l$. 

The effects of uncorrelated fluctuations are shown in the lower part of Fig.~\ref{ps_r6}.
We have plotted there spectra of five simulated distributions.
In the simulations the count rates in the pixels were
completely uncorrelated and the distribution of the count rates in the pixels was Gaussian
with a mean value of $75$ and a dispersion of $25$. As expected, the simulated spectra
are flat. The scatter between the simulations reflecting the stochastic nature of the problem
indicates the level of uncertainties involved in the spectrum estimates.

\begin{figure}
\centering
\includegraphics[width=0.8\linewidth]{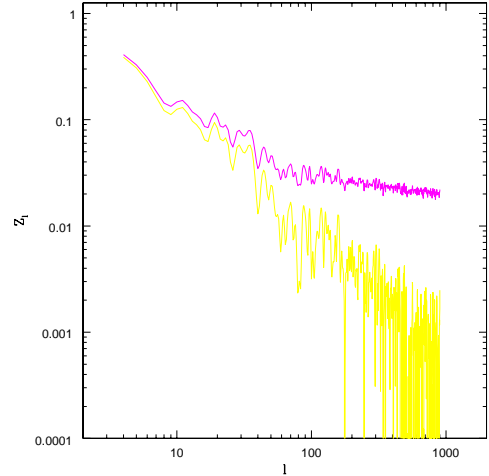}
\caption{An example of the decomposition of the observed spectrum;
upper curve: the power spectrum of the data in area `D' after the removal
of 1\,\% pixels with the highest count rates (same as spectrum No 4 in Fig.~\ref{ps_r6});
lower curve: the residual spectrum obtained from the upper curve by subtraction
of the flat spectrum representing randomly distributed sources contributing most to the
high $l$ part of the total spectrum
(see text).
\label{ps_diff}}
\end{figure}

Before we investigate this question in detail, it is worth  noting that the pixels with
the lowest count rates do not affect significantly the estimates of the power spectrum. 
The low-count-rate part of the distribution is sensitive to errors in
the instrumental background subtraction. As a result of inaccuracies in this
procedure, the XRB signal formally assigned to more than 1000 pixels is negative.
However, negative count rate pixels do not create a long tail in the pixel distribution and
do not influence the overall power spectrum. This is shown
in Fig.~\ref{ps_r6} by spectra No 4 and 5. In both cases an upper count rate
threshold of 256 was used, and in spectrum No 5 all pixels with 
negative count rates have also been removed.

A variety of sources contributing to the XRB produce its complex structure.
Sources which are spatially unrelated (e.g. nearby galaxies
and distant AGNs or extragalactic sources and the local Galactic emission)
create  patterns on the celestial sphere with characteristic and
mutually independent statistical properties. Thus, the power spectrum of
the total XRB signal is a sum of several distinct components corresponding to
separate XRB constituents. The present data do not constrain strongly
amplitudes of individual potential components of the measured spectrum.
To illustrate this we consider spectrum No 4 shown in Fig.~\ref{ps_r6}.
The flat high degree part of this spectrum indicates that a substantial fraction
of the signal in this range of $l$ is produced by small scale fluctuations which
are generated mainly by point-like sources. The distribution of these sources
is not constrained. They may be distributed nonuniformly and contribute also to
the large scale fluctuations described
by the low $l$ section of the spectrum or they may be distributed randomly
so that the
entire signal at small $l$ is due to the other sources contributing to the XRB.
In the framework of the latter model, there are two separate populations contributing
to the XRB. The spectra  of both components have been estimated in the following way.
Using the procedure described in  Sect.~\ref{correct_below}
we calculated the model spectrum assuming that the count rates in the pixels are
completely uncorrelated. This spectrum was then subtracted from the No 4 spectrum; 
the residual spectrum is shown in Fig.~\ref{ps_diff}. The decomposition of the observed
spectrum represents a model in which strong sources produce local fluctuations, while
most of the signal at large angular scales is generated by a smooth distribution. Its
contribution to the local fluctuations is negligible. 

\subsection{Power spectra vs. energy}

\begin{figure}
\centering
\includegraphics[width=0.8\linewidth]{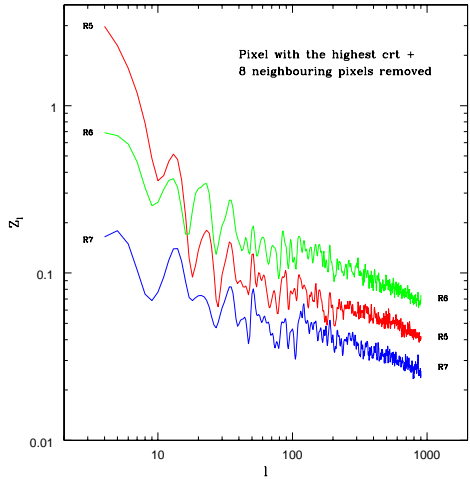}
\caption{The harmonic power spectra of the RASS energy bands R5, R6 and R7; area `D' is used.
One pixel (the same in  all the bands)
with the highest count rate is removed from the data.
\label{r567_1}}
\end{figure}

\begin{figure}
\centering
\includegraphics[width=0.8\linewidth]{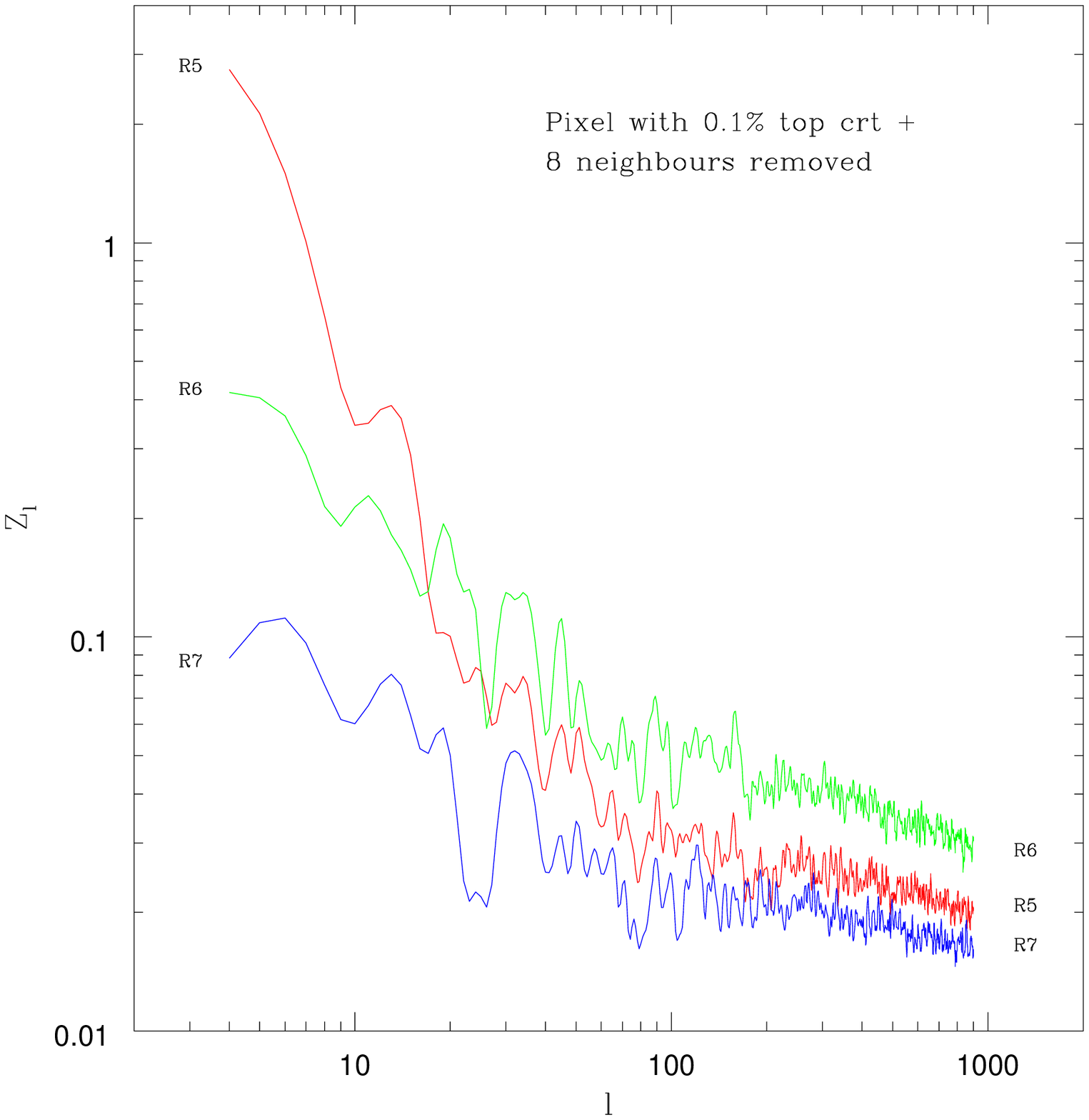}
\caption{Same as Fig.~\ref{r567_1}; the pixels with the top $0.1$\,\% count rates are removed.
\label{r567_2}}
\end{figure}

\begin{figure}
\centering
\includegraphics[width=0.8\linewidth]{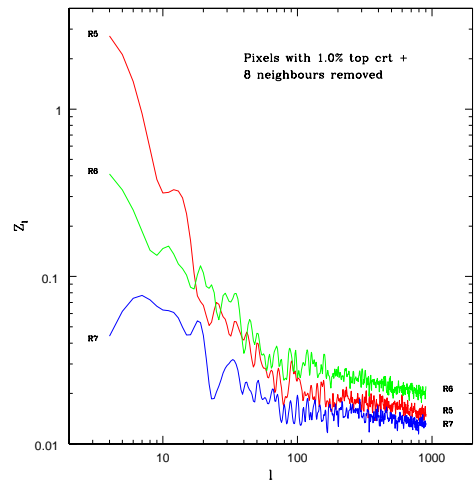}
\caption{Same as Fig.~\ref{r567_2}; the pixels with the top $1.0$\,\% count rates are removed.
\label{r567_3}}
\end{figure}

Because the RASS data useful for the extragalactic analysis cover a narrow range of energies
and the \ROSAT energy bands overlap substantially, the power spectra of R5, R6 and R7
share many common features. Nevertheless they exhibit some differences which can be
ascribed to a slightly different content of the XRB at these energies. In Figs.~\ref{r567_1},
\ref{r567_2} and \ref{r567_3} the power spectra of area `D' of the RASS are shown
in three energy bands. The spectra plotted in Fig.~\ref{r567_1} are calculated after the
removal of the pixel with the highest count rate in the field (produced by
a Seyfert galaxy, IC\,3599);
in Figs.~\ref{r567_2}
and \ref{r567_3} the pixels of the top $0.1$\,\% and $1.0$\,\% count rates were removed.

In each subsample the spectra of all three bands, R5, R6 and R7 exhibit a similar slope
at degrees above  $l \sim 200 - 300$. Also the normalization of all spectra at high $l$ end
is consistent with the conjecture that the amplitudes of the fluctuations at small angular 
scales
are proportional to the average signal in each energy band. This implies that small-scale
fluctuations have the same ``colors'' as the average XRB.

The similarity of the R6 and R7 spectra continues to small degrees $l$ and both these spectra
differ significantly from the spectrum of the softest band R5. The latter spectrum exhibits
a sharp increase below $l \sim 20$ in Figs.~\ref{r567_1} and
$l \sim 100$ in Figs.~\ref{r567_2} and \ref{r567_3}. This is a result of the contamination
of the R5 band by soft photons, which evidently exhibit large-scale non-uniformities.
The spectra of subsamples devoid of the strongest sources reveal a change of slope at quite
small angular scales of a few degrees. Evidently, the excess fluctuations of the soft XRB
over the harder component represent a separate XRB component whose origin deserves
careful examination.

\subsection{The autocorrelation function}
The power spectrum may be used to calculate the autocorrelation function (ACF). In the previous
sections the relationship between those two statistics has been used to elucidate the effects 
of the data binning. Here we employ this relationship (Eq.~(\ref{acffin})) to calculate the ACF
for the whole separation range and to compare it with the ACF determinations
given by \cite{soltan9699}. The relationship between ACF and harmonic power spectrum
is discussed in the Appendix.

\begin{figure}
\centering
\includegraphics[width=0.8\linewidth]{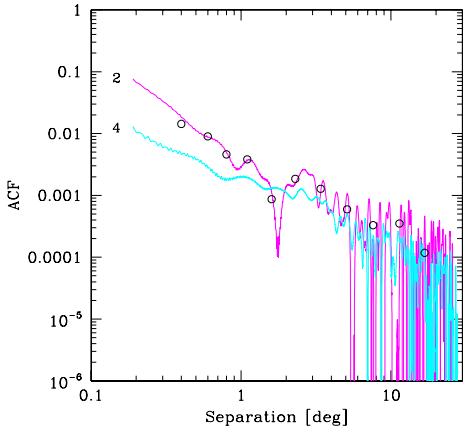}
\caption{The autocorrelations functions of area `D'. Labels `2' and `4' denote
the ACFs computed from the power spectra in Fig.~\ref{ps_r6} marked `2' and `4',
respectively. The ACF calculated directly from the data in subsample No `2' is shown
by open circles.
\label{ps_acf}}
\end{figure}

The ACF computed using two power spectra, No 2 and 4 in Fig.~\ref{ps_r6} is shown in
Fig.~\ref{ps_acf}. For comparison, open circles show the ACF computed directly from
the RASS data using the same subsample as for the curve labelled `2'. It is clearly
visible that the removal of the strongest sources from the area reduces significantly
the ACF at separations below $\sim 1^\circ$. This implies that these sources are not
distributed randomly. The residual ACF signal shows that the XRB devoid of the brightest
sources still exhibits fluctuations. But the substantial decrease of the ACF amplitude
after the removal of $1$\,\% of pixels indicates that the origin of the XRB fluctuations 
requires further studies.

\section{Prospects for the future}

Deciphering the XRB fluctuations plays a crucial role in the investigation of
various components of the background. In the present work we have calculated the
harmonic power spectrum of several subsamples of the RASS. The data have been selected
according to various criteria such as the energy band, amount of the galactic contamination
and  presence of strong sources. The corresponding power spectra reveal the effects of
the different selection criteria. This gives good prospects to use the spectra to investigate 
models of the XRB. 

Such models should specify the relative contributions of various classes of discrete sources
contributing to the XRB as well as the amount of the thermal emission generated by hot
gas which according to \cite{cen99} is accumulating in the gravitational potential
wells of galaxies and galaxy agglomerations. Since the temperature of the gas is relatively
low one expects that the energy spectra of discrete sources and of thermal emission are 
distinctly different. Also a smooth gas distribution would not produce sharp peaks in the XRB. 
These factors could help to differentiate between the two XRB components by means of the power
spectrum and ACF analysis. Such investigation should also utilize the small scale
XRB fluctuation measurements based on the \ROSAT pointings.

\vspace{2mm}
\appendix \noindent
{\bf APPENDIX}\\
The ACF amplitude difference at small separations between ACFs in Fig.~\ref{ps_acf} is in
agreement with the expected correspondence that the larger amplitude of the spectrum at high
degree $l$ generates the stronger signal of the ACF at small separations. However, the
relationship defined by Eq.~(\ref{acffin}) does not yield a plain dependence of the ACF
slope on the slope of the power spectrum $Z_l$. 
Generally the power law shape of the spectrum does not produce the power law shape of the ACF.
However, for the  special case of $Z_l \sim l^{-1}$ Eq.~(\ref{acffin}) generates
an ACF which
is well approximated by a power law with a slope of $-1$. The exact solution differs
from the power law  $w(\theta) \sim 1/\theta$ by less than $5$\,\% for separations
$\theta < 18.\!\!^\circ5$.

One should note that distinctly different harmonic power spectra could produce similar ACFs.
For instance, the ACFs calculated using
spectra plotted in Fig.~\ref{ps_diff} are identical for separations above $\sim 0.\!\!^\circ3$
despite large and systematic differences in the input spectra. This is because the flux
correlations between pixels are the same in both cases. The lower spectrum has been obtained 
from the upper one by subtracting the flat spectrum which represents random (uncorrelated) 
fluctuations.
Thus, both spectra describe distributions with the same correlation properties.

\vspace{2mm}
ACKNOWLEDGEMENTS. The \ROSAT project has been supported by the German Bundesministerium
f\"ur Bildung und Forschung/Deutsches Zentrum f\"ur Luft- und Raumfahrt
(BMBF/DLR) and by the Max-Planck-Gesellschaft (MPG).
This work has been partially supported by the Polish KBN grant 5~P03D~022~20.

\end{document}